\documentclass[a4paper,11pt]{article}

\usepackage{amssymb,amsmath,amsthm,times,mathrsfs,fullpage}
\usepackage{tikz,wrapfig,multirow}
\usepackage[pdftex]{hyperref}
\hypersetup{plainpages=True, pdfstartview=FitH, bookmarksopen=true,
pdfpagemode=UseNone, colorlinks=true,linkcolor=blue,citecolor=blue}
\usepackage{epstopdf}
\usepackage{mathtools}
\usepackage{color}
\usepackage{float}

\allowdisplaybreaks

\newtheorem{thm}{Theorem}

\newtheorem{lmm}{Lemma}

\newtheorem{prop}{Proposition}
\newtheorem{Def}{Definition}
\theoremstyle{remark}
\newtheorem{Rem}{Remark}

\renewcommand{\(}{\left(}
\renewcommand{\)}{\right)}
\newcommand{\Ord}[1]{{\mathcal O}\left(#1\right)}
\newcommand{\rcp}[1]{\frac{1}{#1}}

\newcommand{\bt}{\begin{thm}}
\newcommand{\et}{\end{thm}}
\newcommand{\bl}{\begin{lmm}}
\newcommand{\el}{\end{lmm}}
\newcommand{\bp}{\begin{prop}}
\newcommand{\ep}{\end{prop}}
\newcommand{\bdf}{\begin{Def}}
\newcommand{\edf}{\end{Def}}
\newcommand{\brem}{\begin{Rem}}
\newcommand{\erem}{\end{Rem}}
\newcommand{\bpf}{\begin{proof}}
\newcommand{\epf}{\end{proof}}

\newcommand{\N}{{\mathbb N}}

\newcommand{\C}{{\mathbb C}}

\newcommand{\cd}{\stackrel{d}{\longrightarrow}}

\newcommand{\pdiff}[2]{\frac{\partial^{#2}}{\partial #1^{#2}}}

\newcommand{\E}[1]{\mathbb{E}\left[#1\right]}
\newcommand{\Rm}{R^{[m]}(z)}
\newcommand{\G}[1]{G_{#1}^{[m]}(z)}
\newcommand{\Hm}{H^{[m]}(z)}
\newcommand{\at}[1]{\biggr\rvert_{#1}}

\newcommand{\cA}{\mathcal{A}}
\newcommand{\cB}{\mathcal{B}}
\newcommand{\cC}{\mathcal{C}}
\newcommand{\cD}{\mathcal{D}}

\newcommand{\cR}{\mathcal{R}}

\newcommand{\seq}{\textsc{Seq}}

\newcommand{\ie}{\emph{i.e.}, }
\newcommand{\cf}{\emph{cf.}, }

\begin{document}

\title{\textbf{Sackin Indices for Labeled and Unlabeled Classes \\ of Galled Trees}}
\author{Michael Fuchs\thanks{Partially supported by NSTC under the grant NSTC-113-2115-M-004-004-MY3.} \\
    Department of Mathematical Sciences\\
    National Chengchi University\\
    Taipei 116\\
    Taiwan \and
Bernhard Gittenberger\\
    Institute for Discrete Mathematics and Geometry\\
    TU Wien\\
    1040 Wien\\
    Austria}
\date{February 20, 2025}
\maketitle

\vspace*{-0.5cm}
\begin{center}
{\it Dedicated to Michael Drmota on the occasion of his 60th birthday.}
\end{center}

\begin{abstract}
The Sackin index is an important measure for the balance of phylogenetic trees. We investigate two extensions of the Sackin index to the class of galled trees and two of its subclasses, namely simplex galled trees and normal galled trees, where for all classes we consider both labeled and unlabeled galled trees. In all cases, we show that the mean of the Sackin index for a network which is uniformly sampled from its class is asymptotic to $\mu n^{3/2}$ for an explicit constant $\mu$. In addition, we show that the scaled Sackin index converges weakly and with all its moments to the Airy distribution.
\end{abstract}

\emph{AMS 2020 subject classifications.} 05C20, 60C05, 60F05, 92D15.\\

\emph{Key words.} Galled tree, Sackin index, asymptotic mean, limit law, Airy distribution.

\section{Introduction and Results}

{\it Phylogenetic trees} are used in evolutionary biology to model the ancestor relationship of a set of taxa (\cite{SeSt,St}). In order to judge the suitability of a model, biologists have proposed many different shape parameters which measure how ``balanced" a phylogenetic tree is; see the recent comprehensive survey \cite{FiHeKeKuWi} for a precise definition. One of the oldest of these {\it balance indices} is the Sackin index which is defined as the sum of the root-to-leaves distances over the set of all leaves (again see \cite{FiHeKeKuWi} for a proof that this is indeed a balance index). Statistical properties of the Sackin index have been obtained under several null models. For instance, for the {\it uniform model} or {\it PDA model}, where phylogenetic trees are uniformly sampled, the first-order asymptotics of the mean and variance have been found in \cite{BlFrJa}. Exact results for mean and variance are also known; see \cite{CoMiRoRo,FuJi,KiRo,MiRoRo}. In addition, the authors in \cite{BlFrJa} also showed that the scaled Sackin index converges to the Airy distribution. Similar results were also proved for a closely related random variable, namely, the total path-length of Catalan trees; see, e.g., \cite{FiKa,Ta}.

Despite the popularity of phylogenetic trees, they are often inappropriate as a model for evolution; see Chapter 10 in \cite{St}. In particular, in the presence of reticulation events, phylogenetic trees need to be replaced by {\it phylogenetic networks}. We start with a precise definition of these objects.

\begin{Def}
A (rooted, binary) \textbf{phylogenetic network} is an acyclic, directed graph with no multiple edges whose nodes fall into four categories:
\begin{itemize}
\item[(i)] A unique \textbf{root} which is a node with indegree $0$ and outdegree $2$;
\item[(ii)] \textbf{Leaves} which are nodes with indegree $1$ and outdegree $0$;
\item[(iii)] \textbf{Tree nodes} which are nodes with indegree $1$ and outdegree $2$;
\item[(iv)] \textbf{Reticulation nodes} which are nodes with indegree $2$ and outdegree $1$.
\end{itemize}
\end{Def}
\begin{Rem}
Phylogenetic trees are phylogenetic networks without reticulation nodes.
\end{Rem}
The leaves of a phylogenetic network are usually labeled with the elements from a set $X$ of taxa (where each label is only used once). We then call the phylogenetic network {\it labeled}; otherwise, it is called {\it unlabeled}. Recent years have witnessed a growing body of work on enumeration and stochastic properties of shape parameters of the class of phylogenetic networks and its subclasses when networks are uniformly sampled from a given class; see, e.g., \cite{BoGaMa,FuGiMa,FuYuZh1,FuYuZh2,DiSeSt,Stu}. However, very little is known about the extension of the Sackin index to networks.

First, we need to clarify what we mean by the Sackin index of a phylogenetic network. In fact,
since there may be several paths from the root of the network to a given leaf, different
extensions of the Sackin index from trees to networks are possible. For instance, one can consider
for every leaf the path of maximal length and sum these path lengths over all leaves.
Alternatively, the path of minimal length can be considered. We
investigate both these extensions in this paper and state a formal definition of the Sackin
indices below. Other variants are conceivable, as for each gall (see below for the definition)
that a path must go through, one has to decide which of the two possible ways to go. In the first
extension (maximal length) one always chooses the longer one, in the second the shorter
path is chosen. One may think of variants, e.g., choosing for each gall at random where to go.

The only stochastic result for a Sackin index of a phylogenetic network which we are aware of was obtained in \cite{Zh} where the first extension above was considered for the class of one-component tree-child networks (called simplex networks in \cite{Zh}). More precisely, it was proved that the mean of the Sackin index for a randomly chosen simplex tree-child network with $n$ labeled leaves has the (unusual) order $n^{7/4}$; see \cite{ChFuLiWaYu} for a generalization of this result to $d$-combining simplex tree-child networks. No results for the variance and limit law of a Sackin index for networks have so far been reported in the literature.

The main purpose of this paper is to prove such results for the class of galled trees, which is a popular
network class in phylogenetics; see \cite{HuRuSc,SeSt2}. To define it, we need the notion of a
{\it tree-cycle} or {\it gall} which is a set of two edge-disjoint paths from a tree node to a reticulation node
with all intermediate nodes being also tree nodes.

\begin{Def}
A phylogenetic network is called a \textbf{galled tree} (or \textbf{level-$1$ network}) if (i) every reticulation node is in a tree-cycle and (ii) any two tree-cycles are node-disjoint.
\end{Def}

\begin{Rem}
Note that galled trees are not trees in the classical graph-theoretical sense. Nevertheless, we will refer to them as ``trees" throughout this paper.
\end{Rem}

Galled trees with $n$ labeled leaves have been enumerated exactly in \cite{SeSt2} and asymptotically in \cite{BoGaMa}. Moreover, exact enumeration results for the following two subclasses of labeled galled trees have been obtained in \cite{CaZh}. For the first of these subclasses, a recursion relation and their asymptotic number in the unlabeled case were also recently obtained in \cite{AgMaRo}.

\begin{Def}
A galled tree is called \textbf{simplex} (or one-component) if every reticulation node is followed by a leaf. A galled tree is called \textbf{normal} if the parents of each reticulation node are not in an ancestor-descendant relationship.
\end{Def}
\begin{figure}[H]
\begin{center}\includegraphics[width=12cm]{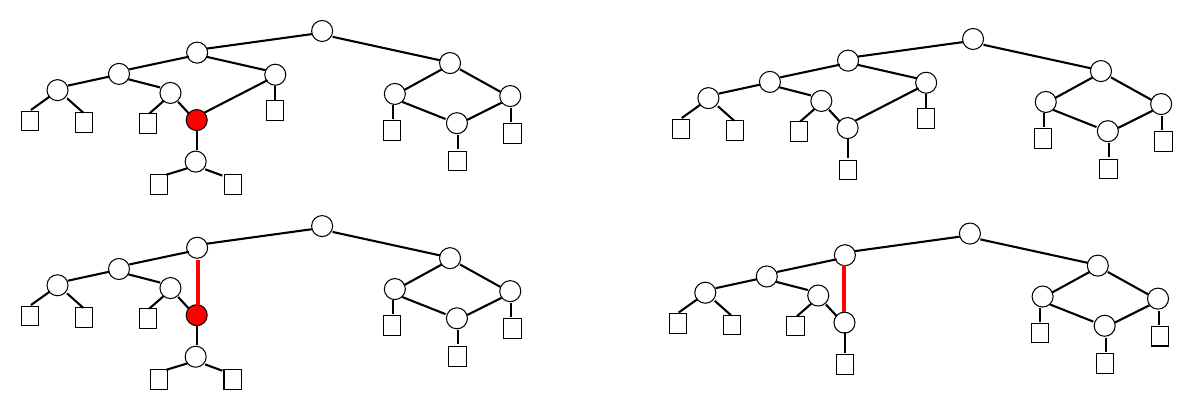}\end{center}
%\vspace*{-5mm}
\caption{\small Four unlabeled galled trees, each with exactly two galls. Upper left: A normal galled tree that is not simplex, as the child of the red reticulation node is not a leaf. Upper right: A simplex and normal galled tree. Lower left: A galled tree that is not normal, because the end points of the red edge are in a ancestor-descendant relationship. It is not simplex either because of the child of the red reticulation node.  Lower right: A simplex, but not normal galled tree.}
\label{galled_types}
\end{figure}

\begin{Rem}\label{one-normal}
Simplex galled trees form an important building block in the construction of networks; see \cite{CaZh,FuYuZh2}. Normal galled trees may be the most relevant galled trees in practical applications because galled trees serve as models for reticulate evolution and the normal ones are exactly the rankable galled trees (\ie they can be generated by an evolution process; see \cite{BiLaSt}).
\end{Rem}

We derive asymptotic counting results for these three classes (galled trees, simplex galled
trees, and normal galled trees) in the next section (both for the labeled and unlabeled
case). Our derivation will use the symbolic method as in \cite{BoGaMa} for labeled galled trees and will
thus be different from the method used in \cite{AgMaRo} for unlabeled normal galled trees; in all other cases, the result will be new. However, our main focus in this paper is on moments and distributional results for
the Sackin index which we formally define next. First, we need the notion of height for a leaf.

\begin{table}[t]
\def\arraystretch{1.8}
\begin{center}
\begin{tabular}{ccc}%\hline
  network class & $(*,**)$ & $\mu^{(*,**)}$ \\ \hline
  galled trees & \begin{tabular}{c} $(\ell,\max)$ \\ $(\ell,\min)$ \end{tabular} & \begin{tabular}{r @{${}\approx{}$} l} \;\:$\frac{47}{32}+\frac{39\sqrt{17}}{544}$ &  $1.764340293\dots$ \\
  \;\:$\frac{49}{32}-\frac{7\sqrt{17}}{544}$ & $1.478195332\dots$ \end{tabular}\\
  \cline{2-3}
  simplex galled trees & \begin{tabular}{c} $(\ell,\max)$ \\
  $(\ell,\min)$ \end{tabular} & \begin{tabular}{r @{${}\approx{}$} l}$\sqrt{(2+\sqrt2)/2}$ & $1.3065629648\dots$ \\
  $\sqrt{(2+\sqrt2)/2}$ & $1.3065629648\dots$ \end{tabular}\\
  \cline{2-3}
  \multirow{2}{*}{normal galled trees} & $(\ell,\max)$ & 
  \hphantom{$\sqrt{(2+\sqrt2)/2}$}
  $\approx 0.6434228878\dots$ \\
  & $(\ell,\min)$ & \hphantom{$\sqrt{(2+\sqrt2)/2}$}
  $\approx 0.6062795760\dots$ \\
  \cline{2-3}
  \hline
\end{tabular}
\end{center}
\caption{\emph{The mean constant for the different classes of galled trees and different Sackin
indices in the labeled cases.}}\label{mean_const}

\bigskip

\def\arraystretch{1.3}
\begin{center}
\begin{tabular}{ccc}%\hline
  network class & $(*,**)$ & $\mu^{(*,**)}$ \\ \hline
  galled trees & \begin{tabular}{c} $(u,\max)$ \\ $(u,\min)$ \end{tabular}& \begin{tabular}{l} $\approx 1.709905157\dots$ \\
  $\approx 1.4350664453\dots$ \end{tabular}\\
    \cline{2-3}
  simplex galled trees & \begin{tabular}{c} $(u,\max)$ \\
  $(u,\min)$ \end{tabular} & \begin{tabular}{l} $\approx 1.2514790858\dots$ \\
  $\approx 1.2514790858\dots$ \end{tabular}\\
  \cline{2-3}
  normal galled trees & \begin{tabular}{c} $(u,\max)$ \\ $(u,\min)$ \end{tabular} & \begin{tabular}{l} $\approx1.125542584\dots$ \\
  $\approx 1.0632588514\dots$ \end{tabular}\\
  \cline{2-3}
  \hline
\end{tabular}
\end{center}
\caption{\emph{The mean constant for the different classes of galled trees and different Sackin
indices in the unlabeled cases.}}\label{mean_const_unlab}
\end{table}

\begin{Def}
Let $N$ be a galled tree and $x$ be one of its leaves. Then the \textbf{longer height} $h^+(x)$ of $x$ is
the length of a longest path from the root of $N$ to $x$. Likewise, we define analogously the
\textbf{shorter height} $h^-(x)$ by the length of a shortest path from the root of $N$ to $x$.
\end{Def}

\begin{Def}
Let $N$ be a galled tree and $L$ be the set of all leaves of $N$. Then, depending on the choice of
the height definition, we define the \textbf{Sackin index} of $N$ in two ways.
\[
S^{+}(N):=\sum_{x\in L} h^+(x), \qquad S^-(N):=\sum_{x\in L} h^-(x).
\]
\end{Def}

We use throughout the work the notation $S_n^{(*,**)}$ where $*\in\{\ell,u\}$ (labeled and unlabeled) and $**\in\{\max,\min\}$
(maximal paths, minimal paths) to denote the Sackin indices of a random galled tree
(or random simplex galled tree or random normal galled tree) with $n$ leaves, where random
means that the galled tree is sampled uniformly at random from its class.

Our first main result concerns the mean of the Sackin indices.
\begin{thm}\label{main-thm-1}
For all cases, as $n\rightarrow\infty$,
\[
{\mathbb E}\left(S_n^{(*,**)}\right)\sim\sqrt{\pi}\mu^{(*,**)} n^{3/2},
\]
where $\mu^{(*,**)}$ with $*\in\{\ell,u\}$ and $**\in\{\max,\min\}$ is given in Table~\ref{mean_const} and Table~\ref{mean_const_unlab}.
\end{thm}

In Tables~\ref{mean_const} and~\ref{mean_const_unlab} we observe that the mean constants are decreasing from general to simplex to normal galled trees. The reason for this is that the number of galls is decreasing in the same way. Its distribution was determined in \cite{BoGaMa} for the general labeled case and in \cite{En} for the other labeled cases. Indeed, it is easily seen that replacing a gall by a binary tree with the same number of leaves diminishes the Sackin index.

Another observation is that simplex galled trees behave in a special way, as their mean constants for both Sackin indices coincide. We only have an intuitive explanation for this, namely that this phenomenon originates from the fact that in simplex galled trees a leaf-to-root path can only pass through a single gall or no gall at all. As the counting sequence of galled trees very much behaves like one of a class of trees, the galls may be seen as perturbations of the underlying tree structure. Such perturbations appear for instance when we blow up a simply generated tree to a P\'olya tree by adding so-called $D$-forests (see \cite{GiJiWa}) and are small, i.e. logarithmic in size. Similarly, in the theory of random graphs as well as in some contexts of machine learning, tree-like graphs are identified by logarithmically sized bi-connected components ({\em cf.} \cite{PaSt} and \cite{LuTrVeZw}). So we expect that the galls are only of size $\log n$ at most. As for a change in the mean constant when switching from $S^+(N)$ to $S^-(N)$ the loss of height of a leaf must be on average of order $\sqrt n$, this is impossible when passing only through at most one gall. 

\medskip
Our second main result generalizes the expansion of the mean from Theorem~\ref{main-thm-1} to all higher moments and thus also gives a limiting distribution result for the Sackin indices.

\begin{thm}\label{main-thm-2}
In all cases, weakly and with convergence of all moments, as $n\rightarrow\infty$,
\[
\frac{S_n^{(*,**)}}{\mu^{(*,**)}}\stackrel{d}{\longrightarrow}S,
\]
where $S$ is the Airy distribution.
\end{thm}
\begin{Rem}
The Airy distribution $S$ is the distribution which is uniquely determined by the following moment sequence:
\[
{\mathbb E}\left(S^m\right)=\frac{2\sqrt{\pi}}{\Gamma((3m-1)/2)}\Omega_m,
\]
where $\Gamma(x)$ is the gamma function and $\Omega_m$ is recursively given by $\Omega_1=1/2$ and
\[
\Omega_m=\frac{m(3m-4)}{2}\Omega_{m-1}+\frac{1}{2}\sum_{\ell=1}^{m-1}\binom{m}{\ell}\Omega_{\ell}\Omega_{m-\ell},\qquad (m\geq 2).
\]
\end{Rem}
\begin{Rem}
For the class of labeled galled trees, the convergence-in-distribution part of Theorem~\ref{main-thm-2} is also a consequence of Theorem 1.2 in \cite{Stu}. However, note that the scaling factor was not explicitly determined in \cite{Stu}.
\end{Rem}

We conclude the introduction with a brief sketch of the paper. In the next section, we recall the symbolic method and explain how it can be used for the enumeration of (labeled and unlabeled) galled trees. In Section~\ref{mean}, we derive the mean of both Sackin indices for labeled galled trees. These results are generalized to all moments in Section~\ref{ll} which then also gives the limit law of the Sackin indices via the method of moments. In Section~\ref{variants}, we consider variants of the class of labeled galled trees. In Section~\ref{unlabeled}, we derive similar results as in Section~\ref{mean}-\ref{variants} for unlabeled galled trees. We conclude the paper with some remarks in Section~\ref{con}.

\section{Symbolic Method and Generating Functions}

We will present what is needed of the symbolic method from analytic combinatorics. The method
starts from combinatorial structures and uses \emph{constructions} to build more complex objects. The
power of the methods comes from a dictionary translating these constructions into algebraic
operations on functions associated with the combinatorial structures, thus making the counting
problems amenable to the huge arsenal of complex analysis. The primary source for this is \cite{FlSe}, where many more constructions are presented along with numerous applications and more advanced
methods.

\subsection{Labeled structures and generating functions}
\label{GF_intro}

We start with the basic definitions.

\bdf
A (labeled) \textbf{combinatorial structure}, which is often also called combinatorial class, is a pair
$(\mathcal A,|\cdot|)$, where $\mathcal A$ is a set of elements, called combinatorial objects,
which are composed of distinguishable (\ie labeled) atoms, and a size function $|\cdot|:\mathcal
A \to \mathbb N_{0}$ such that for all $n\in \N_{0}$ the set $\mathcal A_n:=\{x\in\mathcal A\text{ such
that } |x|=n\}$ is finite.
%\footnote{In this paper, $\mathbb N$ denotes the set of all nonnegative integers.}
%The atoms of every object $x$ carry the labels $1,2\dots,|x|$.

The sequence $(a_n)_{n\in\mathbb N_0}$ with $a_n=\#\mathcal A_n$, the cardinality of $\mathcal A_n$,
is called the \textbf{counting sequence} of $\mathcal A$ and the formal power series
$A(z)=\sum_{n\ge 0} a_n\frac{z^n}{n!}$ is the (exponential) \textbf{generating function}
associated with  $\mathcal A$.
\edf

\brem
If the $(a_n)_{n\in\mathbb N_0}$ does not grow too fast, then $A(z)$ has a positive radius of
convergence and thus represents a function in some neighbourhood of the origin in $\C$.
\erem

\newpage

The above mentioned dictionary translates combinatorial constructions (\ie set operations
involving the ground sets of combinatorial structures with compatible size functions) into
algebraic operations on their generating functions. The concept is usually applied in the
following way: Starting point is a combinatorial counting problem of the form "`How many objects
of size $n$ are there in structure $\cA$?"' The next step is to find a \emph{specification} for
$\cA$ using combinatorial constructions and known combinatorial structures. Finally, apply the
dictionary and analyze the resulting generating functions.

\subsubsection*{Examples for combinatorial constructions}

\begin{itemize}
\item \emph{Combinatorial sum}: Given two combinatorial structures $(\mathcal A,|\cdot|_{\mathcal
A})$ and $(\mathcal B,|\cdot|_{\mathcal B})$, let
$$
\mathcal C=\mathcal A ~\dot\cup~ \mathcal B, \qquad
|x|_{\mathcal C}=\begin{cases}
|x|_{\mathcal A} & \text{ if } x\in\mathcal A, \\
|x|_{\mathcal B} & \text{ if } x\in\mathcal B,
\end{cases}
$$
where $\dot\cup$ is the disjoint union. Thus, $C(z)=A(z)+B(z)$.

\item \emph{Combinatorial product}: Given two combinatorial structures $(\mathcal
A,|\cdot|_{\mathcal A})$ and $(\mathcal B,|\cdot|_{\mathcal B})$ set
$$
\mathcal C=\mathcal A \times \mathcal B, \qquad
|(x,y)|_{\mathcal C}=|x|_{\mathcal A} + |y|_{\mathcal B}.
$$
Every atom of $x$ and every atom of $y$ becomes an atom of $(x,y)$ by this process. In order to
get a correct labeling, relabel $x$ and $y$ in an order-preserving way such that the atoms of
$(x,y)$ carry the labels $1,2,\dots,|x|+|y|$ after all. This relabeling is not unique, as we may
choose which labels go into $x$, leaving the remaining labels for $y$. But as we do respect the
order of the labels within $x$ and within $y$, each such choice results in a unique labeling of
$(x,y)$. Altogether, we obtain
$$
c_n=\sum_{k=0}^n\binom nk a_kb_{n-k},
$$
as for $|(x,y)|=n$ the first component $x$ has size between 0 and $n$ and we have then $\binom nk$
possible choices for the labels of the atoms of $x$ inside $(x,y)$. This relation implies
$C(z)=A(z)B(z).$

\item \emph{Symmetric combinatorial product}: Given a combinatorial structure $(\mathcal
A,|\cdot|_{\mathcal A})$, we may construct the combinatorial product $\cA\times \cA$,
but identify $(x,y)$ and $(y,x)$. Here, we actually construct sets $\{x,y\}$
together with the combinatorial size function $|\{x,y\}|_{\cC}=|x|_{\cA}+|y|_{\cB}$. For stressing
the fact that we have a symmetry in the future formulas, let us write this as $\mathcal C=\mathcal
A \ast \mathcal A$, although a more standard notation would be $\textsc{Set}_2(\cA)$.

Note further, that in a pair $(x,y)$ we can never have $x=y$, as the objects are labeled. Thus, on
the generating function level we get $C(z)=\rcp2A(z)^2$.
%\end{itemize}

%\begin{itemize}
\item \emph{Sequences of positive length}: Let $(\mathcal A,|\cdot|)$ be a combinatorial structure
in which no object has size 0. The sequence construction we will use is defined by
$$
\cC=\seq^+(\cA):=\cA~\dot\cup~(\cA\times\cA)~\dot\cup~(\cA\times\cA\times\cA)~\dot\cup~ \dots.
$$
Now, using the dictionary, we get
$$
C(z)=A(z)+A(z)^2+A(z)^3+\dots=\frac{A(z)}{1-A(z)}.
$$
\end{itemize}

\brem
In the standard literature the sequence construction has nonnegative length, \ie the empty
sequence is also allowed. As the empty sequence constitutes an object of size zero, we would just
have to add 1 to the generating function $C(z)$, giving $C(z)=1/(1-A(z))$ after all. We chose to
exclude the trivial object, as in all our specifications only positive length sequences will occur.
\erem

\subsection{Unlabeled structures}

Without being too formal, for unlabeled structures we simply dispense with the labeling of the
atoms, which then become indistinguishable. Counting is then more complicated, as symmetries
(non-trivial automorphisms) may occur. Nevertheless, for the basic structures that we discussed
above for the labeled world, not much changes. Instead of exponential generating functions
$A(z)=\sum_{n\ge 0} a_n\frac{z^n}{n!}$, we must use ordinary generating functions $A(z)=\sum_{n\ge
0} a_nz^n$. Then, we get the same relations for combinatorial sum, product and the sequence
construction. In the symmetric combinatorial product, we must cope more carefully with symmetries,
as the components of a pair $(x,y)$ need no longer be different. We will not develop the
theory here and just state that the symmetric combinatorial product $\mathcal C=\cA\ast\cA$ has
(ordinary) generating function $C(z)=(A(z)^2+A(z^2))/2$.

\subsection{The generating function for galled trees}

Let us apply the symbolic method to galled trees. We start with a symbolic equation, which is
basically a context-free grammar that specifies galled trees.

\begin{figure}[H]
\begin{center}\includegraphics[width=12cm]{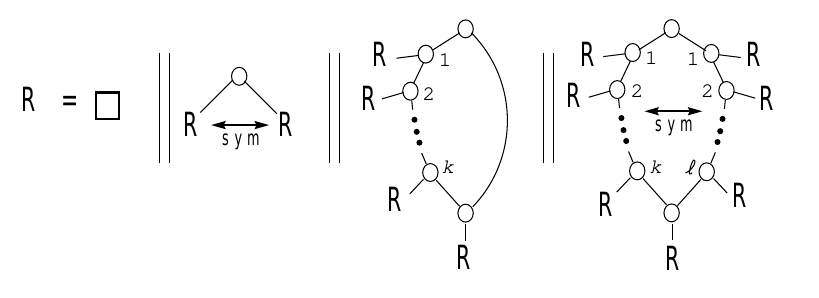}\end{center}
\vspace*{-5mm}
\caption{\small Symbolic specification of galled trees. In the last of these four cases, we assume that the structure is drawn in such a way that $k\ge \ell$.}
\label{R_symb}
\end{figure}

A galled tree can be seen as an object that falls into one of four categories: The first one is a
(labeled) leaf, the simplest possible galled tree. The second consists of a root being a tree
vertex which has two children being galled trees. In this case there is a symmetry, as we may
interchange the two children without changing the galled tree. The third category consists of
galled trees with a root inside the gall. Additionally, we require that there is a nontrivial path
from the root to the (unique) reticulation node of the gall and that the second ingoing edge of
the reticulation node links the root directly to it. The vertices of the nontrivial path are tree
nodes, and their second children are again galled trees. Likewise, the child of the reticulation
node is a galled tree. The last category is similarly shaped, except that instead of the direct
link from the root to the reticulation node there is a nontrivial path of tree nodes.

When regarding the set $\cR$ of galled trees, then the symbolic specification above gives in a
fairly straight-forward way a specification in terms of sets, the ground set of combinatorial
structures.
\begin{equation} \label{R_seteq}
\cR=\{\Box\}~\dot\cup~ \{\circ\}\times\left[(\cR\ast\cR)~\dot\cup~\cR\times \seq^+(\cR)
~\dot\cup~\cR\times (\seq^+(\cR)\ast \seq^+(\cR))\right].
\end{equation}

Now we use our dictionary. This yields a functional equation for the generating function
associated with galled trees:
\begin{equation} \label{R_funeq}
R(z)=z+\rcp2 R(z)^2 +\frac{R(z)^2}{1-R(z)}+ \rcp2 R(z)\(\frac{R(z)}{1-R(z)}\)^2.
\end{equation}

This functional equation is a quartic equation for $R(z)$ and thus has four solutions. Note that $R(z)$
is a generating function, \ie a power series with positive coefficients. Thus, it must be
monotonically increasing and convex on the positive real line near the origin. Out of the four
solutions only one satisfies all conditions. So, the desired generating function is
$$
R(z)=\frac{5-\sqrt{1-8z}-\sqrt{18-8z-2\sqrt{1-8z}}}4= \sum_{n\ge 0} r_n\frac{z^n}{n!}
$$
as has been already shown in \cite{BoGaMa}.

\subsection{Getting the coefficients}

To solve the counting problem, we must compute the coefficients $[z^n]R(z)$ (notation for the
$n$-th coefficient of $R(z)$, \ie, $r_n=n![z^n]R(z)$). The generating function is very explicit and
could be expanded into a Taylor series. We will, however, extract the coefficients asymptotically.
First, often the growth rate is displayed more explicitly by a simple asymptotic formula than with
a complicated exact one. Second, for the study of the Sackin index, we will encounter more involved
generating functions that only allow an asymptotic treatment, so we will need this technique
anyway. This goes back to Flajolet and Odlyzko \cite{FlOd} and is also presented extensively in
\cite{FlSe}.

We start with the necessary definition.
\bdf
A function $f:\cD\to\C$ with $\cD\subseteq \C$ is called $\Delta$\textbf{-analytic} if there are
$\rho,\eta\in\mathbb{R}^{+}$ and $0<\phi<\frac\pi 2$ such that
$f$ is analytic in $\Delta\setminus\{\rho\}$ where
\[
\Delta=\Delta(\eta,\phi)=\{z\,\mid\; |z|\le \rho+\eta, |\arg(z-\rho)|\ge \phi\}.
\vspace*{-4mm}
\]
\begin{figure}[H]%
\begin{center}
\includegraphics[width=30mm]{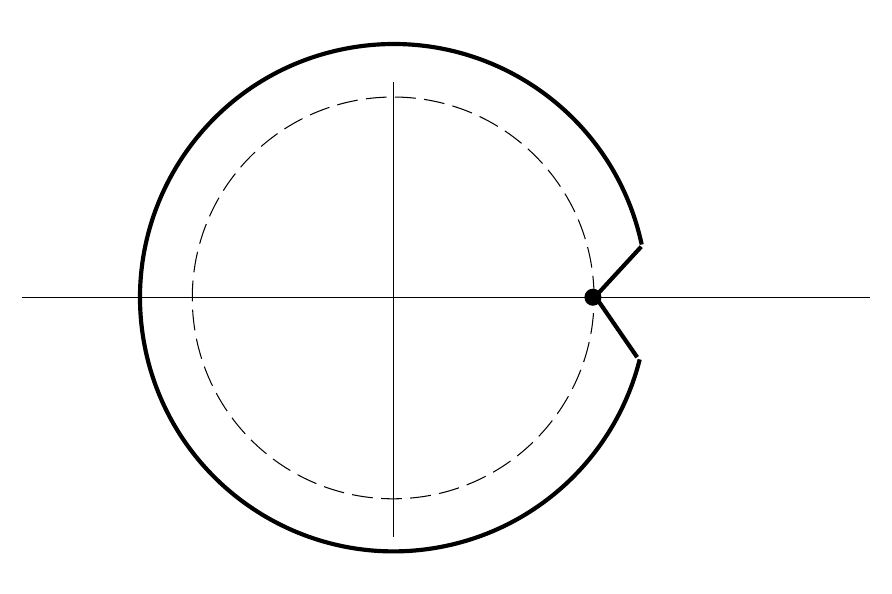}
\vspace*{-4mm}
\end{center}
\caption{\small The domain where we require analyticity.}
\end{figure}
\edf

Now, having a $\Delta$-analytic function that is singular at $\rho$, it turns out that the
behaviour of the function near the singularity $\rho$ determines the behaviour of its
coefficients. In this case, $\rho$ is the closest singularity to the origin. Such singularities
(in principle, there maybe more, leading to more indentations of $\Delta$) are called {\it dominant singularities}. If there is more than one, each contributes to the coefficient asymptotics. Non-dominant singularities do contribute as well, but with smaller exponential order.

\bt[Transfer theorem \cite{FlOd}] \label{TT}
If $f(z)=\sum_n f_nz^n$ is $\Delta$-analytic and
$f(z)\sim \(1-\frac z\rho\)^\alpha$
then
\[
f_n\sim \rho^{-n}\frac{n^{-\alpha-1}}{\Gamma(-\alpha)}.
\]
\et
\begin{Rem}\label{diff-sing-exp}
An important aspect of this result is that it also applies to $f'(z)$. (We will need this below.) More precisely, if $f(z)$ is $\Delta$-analytic and $f(z)\sim\(1-\frac{z}{\rho}\)^\alpha$, then $f'(z)$ is also $\Delta$-analytic with $f'(z)\sim-\frac{\alpha}{\rho}\(1-\frac{z}{\rho}\)^{\alpha-1}$ and thus
\[
[z^n]f'(z)\sim-\alpha\rho^{-n-1}\frac{n^{-\alpha}}{\Gamma(-\alpha+1)}=\rho^{-n-1}\frac{n^{-\alpha}}{\Gamma(-\alpha)};
\]
see \cite{FlOd}.
\end{Rem}

Our function $R(z)$ has a unique dominant singularity at $1/8$, where two of the three radicands
vanish, and satisfies
\begin{align*}
R(z)&=\frac{5-\sqrt{1-8z}-\sqrt{18-8z-2\sqrt{1-8z}}}4 \\
&\sim \frac{5-\sqrt{17}}4-\frac{17-\sqrt{17}}{68}\sqrt{1-8z}, \quad\text{ as }z\to\rcp8.
\end{align*}
Therefore, the transfer theorem above yields (as already computed in \cite{BoGaMa})
\begin{equation} \label{r_n}
r_n\sim \frac{17-\sqrt{17}}{136}\cdot\frac{8^nn!}{\sqrt{\pi n^3}}
 \sim \frac{(17-\sqrt{17})\sqrt2}{136} \(\frac 8e\)^n n^{n-1}.
\end{equation}

In order to deal not only with the size of combinatorial structures but with another parameter as well (like the Sackin index), we use a similar technique but on generating functions in two variables: instead of $R(z)$ we will use $R(z,x)$, as shown in the next section. Partial derivatives occurring there will be denoted by the standard subscript notations: $R_z(z,x)$ and $R_x(z,x)$ denote the partial derivatives of $R(z,x)$ with respect to $z$ and $x$, respectively. 

\section{The Mean of the Sackin Index in Galled Trees}\label{mean}

First, let us fix $S^+$, the sum of the lengths of all maximal leaf-to-root paths in the
network, as the variant of the Sackin index to be considered here. Thus, the induced random
variable associated with networks with $n$ leaves is denoted by $S_n^{(\ell,\max)}$; see
Table~\ref{mean_const}.

In order to study the Sackin index, we need to keep track of it using a second variable $x$ in the
generating function. The variable $z$ keeps track of the size (number of leaves) of the galled
trees. We obtain then a bivariate generating function
$$
R(z,x)=\sum_{n\ge 0}\sum_{k\ge 0} r_{n,k} \frac{z^n}{n!}x^k
$$
with $r_{n,k}$ being the number of galled trees with $n$ leaves and Sackin index equal to $k$.
Given a uniform random network $N$ with $n$ leaves, then the expected Sackin index is
$$
\E{S_n^{(\ell,\max)}} = \frac{\sum_{k\ge 0} kr_{n,k}}{r_n} = \frac{[z^n] R_x(z,1)}{[z^n]R(z)},
$$
and hence is expressible as an $n$-th coefficient of a partial derivative of $R(z,x)$. So, the next
task will be to find a suitable expression for $S(z):=R_x(z,1)$ and then to extract the coefficients.

\medskip
Note that the Sackin index is clearly additive: Indeed, we will see that when going through the
cases listed on the right-hand side of the symbolic equation of Figure~\ref{R_symb}. The first
case, where the network is a single leaf, is trivial. In the second case, each of the two
occurrences of $\mathcal R$ contributes to the Sackin index of the galled tree. Its contribution
is its own Sackin index with the number of its leaves added, as the height of a leaf in the galled
tree is one more than its height in $\mathcal R$. The Sackin index of the galled tree is then the
sum of the contributions of the two subtrees, after all. Likewise in the other cases: each
$\mathcal R$ contributes its Sackin index and its number of leaves multiplied by the height of
the root of $\mathcal R$, as this is the value by which the leaf heights are lifted. Finally, all
contributions are added.

By these additivity properties, our dictionary translating the constructions presented in
Section~\ref{GF_intro} into algebraic operations on generating function remains valid for bivariate
generating functions as well. To cope with the modified leaf heights, note that a term $x^nz^k$ in
the power series $R(z,x)$ represents a network with $n$ leaves and Sackin index $k$. If all leaf
heights are increased by some number, say $m$, then the Sackin index becomes $k+nm$. The galled
tree modified in this way now corresponds to $z^nx^{nm+k}=(zx^m)^nx^k$. We see that in the end we
only have to replace $z$ with $zx^m$, yielding $R(zx^m,x)$. And note further that all we said so
far in this section remains true if we replace $S^+$ with $S^-$.

These observations imply that the same specification as in the univariate case can be used, namely
\eqref{R_seteq} (\emph{cf.} Figure~\ref{R_symb}, where $k$ and $\ell$ denote the lengths of the two paths from the root to the reticulation node of the root gall). And the modifications to the leaf heights caused by the combinatorial constructions
are treated in a straight-forward way. We only have to introduce factors $x^m$ in a suitable way.
This yields
\begin{align}
R(z,x)&=z+\rcp2 R(zx,x)^2 + \sum_{k>0} R(zx^{k+2},x) \prod_{i=1}^k R(zx^{i+1},x)
\nonumber\\
&\qquad + \sum_{\ell\ge 1} \sum_{k>\ell} R(zx^{k+2},x) \prod_{i=1}^k R(zx^{i+1},x)
\prod_{j=1}^\ell R(zx^{j+1},x) \nonumber \\
&\qquad +\rcp 2 \sum_{k\ge 1} R(zx^{k+2},x) \prod_{i=1}^k R(zx^{i+1},x)^2.
\label{biv_funeq_prep}\\
&=z+\rcp2 R(zx,x)^2
+ \sum_{\ell\ge 0} \sum_{k>\ell} R(zx^{k+2},x) \prod_{i=1}^k R(zx^{i+1},x)
\prod_{j=1}^\ell R(zx^{j+1},x) \nonumber \\
&\qquad +\rcp 2 \sum_{k\ge 1} R(zx^{k+2},x) \prod_{i=1}^k R(zx^{i+1},x)^2.
\label{biv_funeq}
\end{align}

Note that the first three terms in \eqref{biv_funeq_prep} match the first three panels in Figure~\ref{R_symb} in their respective order. The double sum in \eqref{biv_funeq_prep} matches the fourth panel in Figure~\ref{R_symb} for the case where $k>\ell$. Finally, the last sum originates from the fourth panel when $k=\ell$. The reason for this splitting of the last case is that (under our assumption $k\geq\ell$) this last subcase is symmetric, whereas the other subcase covered by the fourth panel is not. Indeed, swapping an object from the symmetric case leads to another valid object of the same type and so we must divide by 2 in that case. A slight simplification leads to \eqref{biv_funeq}, after all. 

As mentioned above, we need $S(z):=R_x(z,1)$, so we now differentiate the functional equation
above with respect to $x$ and set $x=1$. Note that $R_z(z,1)=R'(z)$ and that therefore 
\begin{equation}\label{ableitungen}
\left.\pdiff x{} R(zx^i,x)\right|_{x=1}= izR'(z)+S(z). 
\end{equation}
Hence we obtain (omitting the details of some tedious routine calculations that were done with MAPLE)
$
S(z)=zR'(z)f(R(z))+g(R(z))S(z)
$
implying
$$
S(z)=\frac{zR'(z)f(R(z))}{1-g(R(z))},
$$
where
\begin{align}
f(t)&=t+\sum_{\ell\ge 0}\sum_{k>\ell}
\(\binom{k+3}2+\binom{\ell+2}2-2\)t^{k+\ell}\nonumber\\
&\qquad +\sum_{k\ge 1}\(\frac{k+2}2+\binom{k+2}2-1\)t^{2k} \nonumber\\
&=\frac{t(2t^5-8t^4+10t^3-t^2-11t+12)}{2(1-t)^4(1+t)} \label{f}
\end{align}
and 
\begin{align*}
g(t)&=t+\sum_{\ell\ge 0}\sum_{k>\ell}(k+\ell+1)t^{k+\ell}+
\sum_{k\ge 1}\(\frac12+k\)t^{2k} \\
&=\frac{t(-2t^3 + 7t^2 - 9t + 6)}{2(1-t)^3}. 
\end{align*}

As $S(z)$ depends only on $R(z)$ (via $f(t)$ and $g(t)$) and $1-g(R(z))\neq 0$ for $0\le z< 1/8$,
there will be no pole in the denominator. Therefore, the dominant singularity of $S(z)$ is at
$z=1/8$ as well. We can use the singular behaviour of $R(z)$ near $z=1/8$ to determine the
behaviour of $S(z)$. Again, we omit tedious routine calculations and get
$$
%S(z)\sim \frac{95-\sqrt{17}}{544}\cdot\rcp{1-8z}
S(z)\sim \frac{95-\sqrt{17}}{544}\cdot\rcp{1-8z}
\quad\text{ and so }\quad
[z^n] S(z)\sim \frac{95-\sqrt{17}}{544}\cdot 8^n,
$$
and consequently, by \eqref{r_n}, we get 
\[
\E{S_n^{(\ell,\max)}}=
\frac{[z^n]S(z)}{[z^n]R(z)}\sim \(\frac{47}{32}+\frac{39\sqrt{17}}{544}\) \sqrt\pi n^{3/2} \approx
1.764340293\dots{\cdot} \sqrt\pi n^{3/2}.
\]

If we look at $S^-$ instead of $S^+$, then the formal procedure is exactly the same 
with only a little modification on the increments of the leaf height in the 
substructures. In fact, in Figure~\ref{R_symb} we choose the shortest instead of the 
longest path, meaning that we take the single edge in the third case and the $\ell$-
path in the fourth case. This corresponds to changing the exponent $k$ to $\ell$ in 
the double sum of \eqref{biv_funeq}, where the special case $\ell=0$ matches the 
third panel in Figure~\ref{R_symb} and $\ell>0$ corresponds to the asymmetric 
subcase of the fourth panel. This yields therefore
\begin{align*}
R(z,x)&=z+\rcp2 R(zx,x)^2
 + \sum_{\ell\ge 0} \sum_{k>\ell}R(zx^{\ell+2},x)  \prod_{i=1}^k R(zx^{i+1},x)
\prod_{j=1}^\ell R(zx^{j+1},x)\\
&\qquad +\rcp 2 \sum_{k\ge 1} R(zx^{k+2},x) \prod_{i=1}^k R(zx^{i+1},x)^2.
\end{align*}
Differentiating and setting $x=1$ gives this time
$$
S(z)=\frac{zR'(z)f(R(z))}{1-g(R(z))}
$$
with
\begin{align*}
f(t)&=\frac{t(2t^5-8t^4+10t^3-t^2-9t+10)}{2(1-t)^4(1+t)},\\
g(t)&=\frac{t(-2t^3 + 7t^2 - 9t + 6)}{2(1-t)^3},
\end{align*}
and so
$$
%S(z)\sim \frac{105-7\sqrt{17}}{544}\cdot\rcp{1-8z}
S(z)\sim \frac{105-7\sqrt{17}}{544}\cdot\rcp{1-8z}
\quad\text{ and so }\quad
[z^n] S(z)\sim \frac{105-7\sqrt{17}}{544}\cdot 8^n,
$$
and this in conjunction with \eqref{r_n} yields
$$
\E{S_n^{(\ell,\min)}}\sim \(\frac{49}{32}-\frac{7\sqrt{17}}{544}\) \sqrt\pi n^{3/2} \approx
1.478195332\dots{\cdot} \sqrt\pi n^{3/2}.
$$

\section{Limit Law of Sackin Indices}\label{ll}

Our next goal is to find the limit law of the Sackin index of a uniform random galled tree with
$n$ leaves, when $n$ tends to infinity. A weak limit theorem has been shown in \cite{Stu}, even in
a more general setting. However, the result in \cite{Stu} does not give explicit asymptotics of the moments. And in general, a weak limit does not even imply convergence of moments. In order to describe the
limit law, we will compute the asymptotics of all moments and then apply the method of moments,
showing that the limiting distribution is uniquely determined by its moments and implying a weak
limit law as well.

We start with the functional equation \eqref{biv_funeq},
\begin{align}
R(z,x)&=z+\rcp2 R(zx,x)^2 %+ \sum_{k\ge 1} R(zx^{k+2},x) \prod_{i=1}^k R(zx^{i+1},x)
 + \sum_{\ell\ge 0} \sum_{k>\ell} R(zx^{k+2},x)\prod_{i=1}^{k} R(zx^{i+1},x)
\prod_{j=1}^\ell R(zx^{j+1},x) \nonumber\\
&\qquad +\rcp 2 \sum_{k\ge 1} R(zx^{k+2},x) \prod_{i=1}^k R(zx^{i+1},x)^2,
\label{biv_funeq_mod}
\end{align}
and set
\begin{align*}
\Phi(z,A)&=z+\rcp2A^2+\sum_{\ell\ge 0} \sum_{k>\ell} A^{\ell+k+1}+\rcp2\sum_{k\ge1}A^{2k+1}\\
&=z+\frac{1}{2}A^2+\frac{A^2}{(A-1)^2(A+1)}-\frac{A^3}{2(A^2-1)}.
\end{align*}
This function captures the underlying structure of the functional equation: Indeed, if we replace
all the $A$'s by $R(z,1)=R(z)$, we get with $A=\Phi(z,A)$ the functional equation \eqref{R_funeq}
for $R(z)$.

Set $\rho=1/8$. Then, we know already that, as $z\to \rho$,
\begin{equation} \label{R_asym}
R(z,1)\sim \tau-a \sqrt{1-\frac z\rho}
\end{equation}
with
$$
\tau=\frac{5-\sqrt{17}}4 \quad  \text{ and } \quad~ a=\frac{17-\sqrt{17}}{68}.
$$
The $m$-th factorial moment of $S_n^{(\ell,\max)}$ in a uniform random galled tree with $n$ leaves
is related to $[z^n]\Rm$ where
$$
\Rm :=\frac{\partial^m}{\partial x^m} R(z,x)\biggr\rvert_{x=1}.
$$

To get hands on $\Rm$, we need to do an $m$-fold differentiation on \eqref{biv_funeq_mod}. This may
look scary, but we only need the asymptotic main term of $\Rm$ which is given in the next proposition whose proof uses induction starting from (\ref{R_asym}) (and thus gives a second way of deriving the result of the previous section).

%The following lemma will simplify
%our computations considerably.
%
%\bl
%\label{aux_lem}
%For all nonnegative integers $j$ we have
%$$
%\pdiff xm R(zx^j,x)\at{x=1}=\Rm+mjz\frac{\partial^m}{\partial x^{m-1}\partial
%z}R(z,x)\at{x=1}+\sum_{r=2}^m K_r(z)\frac{\partial^m}{\partial x^{m-r}\partial
%z^r}R(z,x)\at{x=1}.
%$$
%\el

\bp
\label{Rm_asym}
For $m\ge1$,
$$
\Rm\sim c_m\(1-\frac z\rho\)^{-(3m-1)/2},
$$
where $c_1=f(\tau)/(2\Phi_{AA}(\rho,\tau))$, with $f$ from \eqref{f}, and
\begin{equation} \label{cm}
c_m=\frac{f(\tau)}{2a\,\Phi_{AA}(\rho,\tau)}m(3m-4)c_{m-1}
+\rcp {2a}\sum_{\ell=1}^{m-1}\binom{m}{\ell}c_{\ell}c_{m-\ell}
\end{equation}
for $m\ge 2$.
\ep

\brem
We will need the moments for the limit theorem, not the factorial moments. The $m$-th moment, however, is a linear combination of all factorial moments up to order $m$. The transfer theorem, Theorem~\ref{TT}, tells us that the asymptotic growth of the coefficients of $\Rm$ depends only on its behaviour when $z\to\rho$. Therefore, Proposition~\ref{Rm_asym} implies that the $m$-th factorial moment and the $m$-th "`ordinary"' moment are asymptotically equal.
\erem
\bpf
We differentiate the functional equation \eqref{biv_funeq_mod} $m$ times with respect to $x$ and set $x=1$. This gives
\begin{equation} \label{Rm_decomp}
\Rm=\G1+\G2+\G3,
\end{equation}
where
\begin{align*}
\G1&=\rcp2\sum_{r_1+r_2=m}\binom{m}{r_1} \pdiff{x}{r_1} R(zx,x)\at{x=1}\pdiff{x}{r_2} R(zx,x)\at{x=1}, \\
\G2&=\sum_{\ell\ge 0}\sum_{k>\ell}\sum_{r_1+\cdots+r_{\ell+k+1}=m} \binom{m}{r_1,\dots,r_{k+\ell+1}}
\prod_{i=2}^{k+2}  \pdiff x{r_{i-1}} R(zx^{i},z) \at{x=1}
\\
&\qquad\times \prod_{j=2}^{\ell+1}\pdiff x{r_{k+j}} R(zx^{j},z)\at{x=1},
\\
\G3&=\rcp2 \sum_{k\geq 1}\sum_{r_1+\cdots+r_{2k+1}=m} \binom{m}{r_1,\dots,r_{2k+1}}
\pdiff x{r_1} R(zx^{k+2},x)\at{x=1}\\
&\qquad\times
\prod_{i=2}^{k+1}
\(\pdiff x{r_{2i-2}} R(zx^{i}\!,z)\at{x=1} \pdiff x{r_{2i-1}} R(zx^{i}\!,z)\at{x=1}\).
\end{align*}
The right-hand side of \eqref{Rm_decomp} contains terms of the form $\pdiff xm R(zx^i,x)$, but by
the multi-dimensional version of the Fa\`{a} di Bruno formula (see e.g. \cite{HEMM}), we get
\begin{equation} \label{Leibniz}
\pdiff xm R(zx^i,x)\at{x=1}=\Rm+miz\frac{\partial}{\partial z}R^{[m-1]}(z)
+\sum_{r_1=0}^{m-2}\sum_{r_2=1}^{m-r_1} K_{r_1,r_2}(z)\frac{\partial^{r_2}}{\partial
z^{r_2}}R^{[r_1]}(z),
\end{equation}
where $K_{r_1,r_2}(z)$ are polynomials in $z$ coming from inner derivatives. Their coefficients depend on $m$ and $i$.

Having this, we can now collect all occurrences of $\Rm$ on the right-hand side of
\eqref{Rm_decomp}, which yields $\Rm=\Phi_A(z,R(z)) \Rm + \Hm$ after all. Solving for $\Rm$ gives
\begin{equation}
\label{Hm}
\Rm=\frac{\Hm}{1-\Phi_A(z,R(z))},
\end{equation}
where $\Hm$ contains terms involving derivatives of $R^{[r]}(z)$ with $r<m$ and
hence the induction hypothesis can be applied. The next goal is therefore to find the terms in
$\Hm$ being asymptotically largest. Equivalently, we search for the terms of order
$(1-\frac z\rho)^{-\alpha}$ with largest $\alpha$.

To this end, note that according to the induction hypothesis, the $\alpha$ of $R^{[r]}(z)$ increases by $3/2$ when $r$ increases by one and increases by $1$ each time $R^{[r]}(z)$ is differentiated; see Remark~\ref{diff-sing-exp}. Thus, the largest $\alpha$ in $\Hm$ is obtained when using one of the two choices in the innermost sum of $\G1$, $\G2$, $\G3$:
\begin{enumerate}
\item Either set all but one of the $r_i$'s equal to zero and pick the second term in the expansion \eqref{Leibniz} of the one term that is differentiated at least once,
\item or set all but exactly two of the $r_i$'s equal to zero and for both of the two
corresponding terms pick the main term of their respective expansions \eqref{Leibniz}.
\end{enumerate}
Collecting all this gives
$$
\Hm\sim d_m\(1-\frac z\rho\)^{-(3m-2)/2},
$$
where $d_1=ab/2$ and for $m\geq 2$
\begin{equation}\label{d_m}
d_m=b\frac{m(3m-4)}2c_{m-1}+\frac{\Phi_{AA}(\rho,\tau)}2\sum_{\ell=1}^{m-1}\binom{m}{\ell}c_{\ell}c_{m-\ell}
\end{equation}
with
\[
b=\frac{\tau(2\tau^5-8\tau^4+10\tau^3-\tau^2-11\tau+12)}{2(1-\tau)^4(1+\tau)}=f(\tau).
\]
\brem
It is not a coincidence that we obtain the same function $f$ as in \eqref{f}, as the first term of \eqref{d_m} originates from the first choice above where the second term in \eqref{Leibniz} is taken. This term introduces a factor $i$, just like the contribution of the differentiation \eqref{ableitungen} to $f$ in the computation \eqref{f}. Thus, we actually perform the same summation as in the computation of $f$ in \eqref{f}. 
\erem

So, we have so far
\[
\Hm\sim d_m\(1-\frac z\rho\)^{-(3m-2)/2} \text{ and }\quad \Rm=\frac{\Hm}{1-\Phi_A(z,R(z))}.
\]
Moreover, note that
\begin{equation} \label{funsys}
\tau=\Phi(\rho,\tau) \quad \text{ and }\quad 1=\Phi_A(\rho,\tau).
\end{equation}
Indeed, this is a consequence of the fact that $R(z)=\Phi(z,R(z))$, $\rho$ is a singular point
of $R(z)$ and $R(\rho)=\tau$, and the implicit function theorem.

From \eqref{funsys}, we get by Taylor's theorem
$$
1-\Phi_A(z,R(z))\sim a\,\Phi_{AA}(\rho,\tau)\sqrt{1-\frac z\rho},
$$
and thus
$$
\Rm\sim c_m\(1-\frac z\rho\)^{-(3m-1)/2}
$$
with $c_m$ as claimed in \eqref{cm}.
\epf

From the last result, we deduce our second main result; compare with Theorem~\ref{main-thm-2}.

\bp
Let $\mu:=f(\tau)/(a\,\Phi_{AA}(\rho,\tau))$. Then,
$$
\frac{S_n^{(\ell,\max)}}{\mu n^{3/2}} \cd  S,
$$
where the law of $S$ is the Airy distribution. In addition, all moments
converge as well.
\ep

\bpf
Apply the transfer theorem, Theorem~\ref{TT}, to the result of Proposition~\ref{Rm_asym}. We get
$$
\E{\(\frac{S_n^{(\ell,\max)}}{\mu n^{3/2}}\)^m}\sim\rcp{\mu^m n^{3m/2}} \frac{[z^n]\Rm}{[z^n]R(z)}\sim
\frac{2\sqrt\pi}{\Gamma\(\frac{3m-1}2\)}\cdot \frac{c_m}{a\mu^m},
$$
where the first asymptotic equivalence follows from the fact that the moments are asymptotic to the factorial moments (see above).

Set $\Omega_m:=c_m/(a\mu^m)$. Then, the recurrence \eqref{cm} for $c_m$ becomes
$$
\Omega_m=\frac{m(3m-4)}2\Omega_{m-1}+\rcp2\sum_{\ell=1}^{m-1}\binom m\ell \Omega_\ell\Omega_{m-\ell}
$$
and $\Omega_1=1/2$. Since $2\sqrt\pi\Omega_m/\Gamma((3m-1)/2)$ are the moments of the Airy distribution (see \cite{FlLo}) and the Airy distribution is uniquely determined by its moments,
the claimed result follows.
\epf

\brem
Using similar computations, we can show that the analogous result holds when $S_n^{(\ell,\max)}$ is
replaced by $S_n^{(\ell,\min)}$.
\erem

\section{Variations}\label{variants}

In this section, we consider the two variants of galled trees introduced in the introduction (simplex galled trees and normal galled trees). Since only Theorem~\ref{main-thm-1} is different for these two variants whereas Theorem~\ref{main-thm-2} remains the same, we will focus on the former; the latter result is proved with the arguments from Section~\ref{ll} with only minor modifications.

\subsection{Simplex galled trees}

\begin{figure}[h]
\begin{center}
\includegraphics[width=10cm]{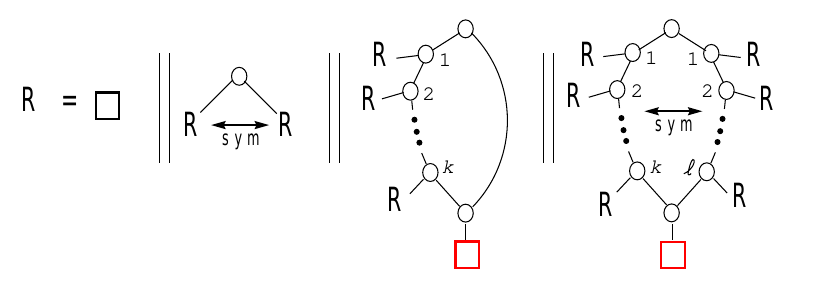}
\end{center}
\vspace*{-5mm}
\caption{\small Symbolic equation for simplex galled trees. The red substructures are the places where
galled trees and simplex galled trees differ.}\label{simplex-spec}
\label{simplex_symb}
\end{figure}

Simplex galled trees, also called one-component galled trees, are another model of galled trees
studied in the literature. They differ slightly from galled trees by imposing the following
constraint: The child of a reticulation node is always a leaf. So, adapting the specification of
galled trees (Figure~\ref{R_symb}) is immediate; see Figure~\ref{simplex-spec}.

The specification in terms of combinatorial structures and constructions is thus given by
\begin{equation} \label{simplex_spec}
\cR=\{\Box\}~\dot\cup~ \{\circ\}\times\left[(\cR\ast\cR)~\dot\cup~\{\Box\}\times\seq^+(\cR)~\dot\cup
~\{\Box\}\times (\seq^+(\cR)\ast \seq^+(\cR))\right].
\end{equation}
Using our dictionary, we obtain the functional equation for the generating function, which is
\begin{equation} \label{simplex_funeq}
R(z)=z+\rcp2 R(z)^2 +\frac{zR(z)}{1-R(z)}+ \rcp2 z\(\frac{R(z)}{1-R(z)}\)^2,
\end{equation}
and admits the exact solution
$$
R(z)=1-\rcp2\sqrt{2-2z+2\sqrt{1-6z+z^2}}
$$
which is the unique power series solution with positive coefficients.

It is easy to see that $R(z)$ is $\Delta$-analytic and has a unique dominant singularity at
$\rho=3-2\sqrt2.$ Near $\rho$, we have the singular expansion
\begin{equation} \label{R_expansion_simplex}
R(z)\sim 1-\sqrt{\sqrt{2}-1}-\frac{\sqrt{2-\sqrt2}}{2}\sqrt{1-\frac{z}{3-2\sqrt2}},
\quad \text{ as } z\to \rho.
\end{equation}
The transfer theorem gives therefore
$$
\frac{r_n}{n!}=[z^n]R(z)\sim \frac{\sqrt{2-\sqrt2}}{4\sqrt{\pi n^3}}(3+2\sqrt2)^n.
$$
Note here that $3+2\sqrt2$ is the multiplicative inverse of $3-2\sqrt2$.

As in the case of galled trees, we use the specification from Figure~\ref{simplex_symb} to derive
a functional equation for the bivariate generating function of $S_{n}^{(\ell,max)}$. When writing \eqref{simplex_spec} in
the form
\begin{align*}
R(z)&=z+\rcp2 R(z)^2 +\sum_{\ell\ge 0} \sum_{k>\ell} z\prod_{i=1}^k R(z)\prod_{j=1}^\ell R(z)+\rcp 2 \sum_{k\ge 1} z \prod_{i=1}^k R(z)^2,
\end{align*}
then each term $R(z)$ corresponds to one of the $\cR$'s in \eqref{simplex_spec}, \cf also
Figure~\ref{simplex_symb}. Therefore, each $R(z)$ must be replaced by a suitable $R(zx^{i+1},x)$.
This gives
\begin{align}
R(z,x)&=z+\rcp2 R(zx,x)^2 + \sum_{\ell\ge 0} \sum_{k>\ell} zx^{k+2} \prod_{i=1}^k R(zx^{i+1},x) \label{simplex_biv}
\prod_{j=1}^\ell R(zx^{j+1},x)
\nonumber \\
&\qquad +\rcp 2 \sum_{k\ge 1} zx^{k+2} \prod_{i=1}^k R(zx^{i+1},x)^2.
\end{align}
Again, we set $S(z)=R_x(z,1)$ and derive the functional equation with respect to $x$ and set
$x=1$. We get
\begin{equation} \label{S_expr}
S(z)=\frac{zR'(z)f(z,R(z))+h(z,R(z))}{1-g(z,R(z))}
\end{equation}
with
\begin{equation} \label{fgh}
f(z,t)=t+z\frac{(2-t)}{(1-t)^4}, \qquad g(z,t)=t+\frac{z}{(1-t)^3},
\qquad
h(z,t)=z\frac{t(2t^3-4t^2-t+6)}{2(1-t)^3(1+t)}.
\end{equation}
From the singular behaviour of $R(z)$ and \eqref{S_expr} we obtain
$$
S(z)\sim \rcp4\cdot\rcp{1-(3+2\sqrt2)z}, \quad\text{ as } z\to 3-2\sqrt2.
$$
This implies
$$
\E{S_n^{(\ell,\max)}}\sim \frac{\sqrt{2+\sqrt2}}{\sqrt2} \sqrt\pi n^{3/2}\approx
1.306563\dots{\cdot} \sqrt{\pi}
n^{3/2}.
%%% \sqrt(pi) noch rausrechnen aus der Numerik.
$$

The functional equation for the bivariate generating function for $S_n^{(\ell,\min)}$ is only
slightly different from the one for $S_n^{(\ell,\max)}$. The only difference is in the double
sum in \eqref{simplex_biv}, where the first factor $zx^{k+2}$ is changed to $zx^{\ell+2}$.
Likewise, $S(z)$ can be expressed in term of $f(z,t)$, $g(z,t)$, and $h(z,t)$ (see \eqref{S_expr}) with
$f(z,t)$ and $g(z,t)$ as in \eqref{fgh} and in $h(z,t)$ only the constant 6 in the numerator is changed
to 4. The asymptotics of $S(z)$ and hence for $\E{S_n^{(\ell,\min)}}$ does not change. Of course,
this affects only the asymptotic main term. If we expand further to the next-order term, then we
observe a difference. For displaying more details, let us write $S^+(z)$ and $S^-(z)$ for the
generating functions of these two cases. And note that the second-order singular term of $R(z)$
is of order $(1-(3+2\sqrt2)z)^{3/2}$. Hence, the term in the asymptotic expansion of $r_n$ is by
a factor $n$ smaller than the main term, whereas it will turn out that the second-order terms of
$[z^n]S^+(z)$ and $[z^n]S^-(z)$ are dominated by the main term only by a factor $\sqrt n$.
Therefore, it suffices to expand $S^+(z)$ and $S^-(z)$, there is no need to look at further terms
of $r_n$. Note, however, that for the further expansion of $S^+(z)$ and $S^-(z)$,
\eqref{R_expansion_simplex} is not enough, \ie we do need the next order term of $R(z)$. But as we
have an explicit expression for $R(z)$, this can be done automatically with MAPLE.

Expanding $S^+(z)$ and $S^-(z)$ up to their second-order terms gives
\begin{align*}
S^+(z)&=\rcp 4\rcp{1-(3+2\sqrt2)z} + \frac{(22+9\sqrt2-(12+76\sqrt2)\sqrt{\sqrt2-1})\sqrt{3\sqrt2-4}}{184 \sqrt{1-(3+2\sqrt2)z}}+\Ord1,
\\[5pt]
S^-(z)&=\rcp 4\rcp{1-(3+2\sqrt2)z} + \frac{(-22-9\sqrt2+(12-62)\sqrt{\sqrt2-1})\sqrt{3\sqrt2-4}}{184\sqrt{1-(3+2\sqrt2)z}}+\Ord1.
\end{align*}
The transfer theorem then implies
\begin{align*}
\E{S_n^{(\ell,\max)}}&=\frac{\sqrt{2+\sqrt2}}{2} \sqrt\pi n^{3/2} +
\frac{(22+9\sqrt2)\sqrt{\sqrt2-1}-140+64\sqrt2}{46} n+\Ord{\sqrt n} \\[4pt]
&\approx 1.306563\dots n^{3/2}-0.589992\dots \sqrt n+\Ord 1, \\[12pt]
\E{S_n^{(\ell,\min)}}&=\frac{\sqrt{2+\sqrt2}}{2} \sqrt\pi n^{3/2} +
\frac{-(22+9\sqrt2)\sqrt{\sqrt2-1}-136+74\sqrt2)}{46} n+\Ord{\sqrt n} \\[4pt]
&\approx 1.306563\dots n^{3/2} -1.167367\dots \sqrt n+\Ord 1.
\end{align*}

\subsection{Normal galled trees}

\begin{figure}[H]
\begin{center}
\includegraphics[width=7cm]{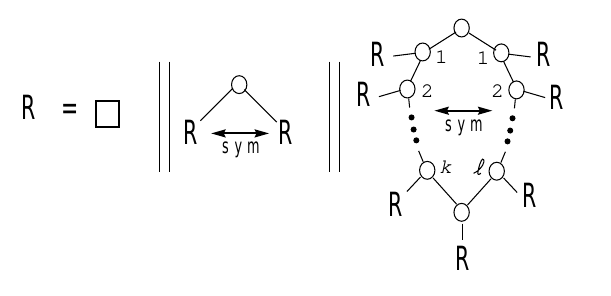}
\end{center}
\vspace*{-0.7cm}\caption{\small Symbolic specification of normal galled trees.}
\label{normal_symb}
\end{figure}

Normal galled trees are precisely the rankable galled trees and therefore of particular practical relevance (see Remark~\ref{one-normal}). Starting again from
Figure~\ref{R_symb}, in the third case the parent nodes of the reticulation node are in an
ancestor-descendant relationship, which is forbidden in normal networks. The other cases do not
violate the normality condition. Therefore, we exclude the bad case and get the symbolic
description shown in Figure~\ref{normal_symb}.

This readily gives the specification in terms of combinatorial structures and constructions,
namely,
\begin{equation} \label{normal_spec}
\cR=\{\Box\}~\dot\cup~ \{\circ\}\times\left[(\cR\ast\cR)~\dot\cup~
\cR\times (\seq^+(\cR)\ast \seq^+(\cR))\right],
\end{equation}
and our dictionary then yields the functional equation for the generating function:
$$
R(z)=z+\rcp2 R(z)^2 + \rcp2 R(z)\(\frac{R(z)}{1-R(z)}\)^2.
$$
Out of the four solutions, exactly one is a power series with positive coefficients. This is
%\begin{align*}
\[R(z)=\frac{3}{4}-\frac{\sqrt{3A}}{12}
%\\
%&\quad
-\rcp3 \cdot\sqrt{\frac{3 \(
3\sqrt3B^{1/3} \(\frac{17}{8}-z\)
-
\(
\frac{B^{2/3}}{4}
+\(2z+\frac{13}{8}\) B^{1/3}+z^{2}+2z+\frac{7}{4}\)
\sqrt{A}
\)}{B^{1/3} \sqrt{A}}},
\]
%\end{align*}
where
\begin{align*}
A&=\frac{4 B^{2/3}-(16z+13)
B^{1/3}+16z^2+32z+28}{B^{1/3}},  \\
B&=8z^3+24z^2-75z+44+3\sqrt{-192z^4-528z^3+645z^2-864z+177}.
\end{align*}
As this is a very involved expression, we proceed by means of the implicit function theorem. Note
that the pair $(\rho,\tau):=(\rho,R(\rho))$, where $\rho$ is the dominant singularity of $R(z)$, is
the solution of the system
\begin{align} \label{sing_sys}
\tau&=\rho+\frac{\tau^2}2+\frac{\tau}2 \(\frac{\tau}{1 - \tau}\)^2,\nonumber\\
1&=\tau+\frac12\(\frac{\tau}{1-\tau}\)^2+\frac{\tau^2}{(1-\tau)^3}.
\end{align}
The second equation of this system is the derivative of the first one with respect to $\tau$.
If we set
\[
F(z,R)=-R+z+\frac{R^2}2+\frac R2\(\frac{R}{1-R}\)^2,
\]
then the system becomes $F(\rho,\tau)=0$, $F_R(\rho,\tau)=0$, which readily yields the singular points of
$R(z)$. Expanding $F$ into a Taylor series around $(\rho,\tau)$ and using $F(z,R(z))=0$ as well
as the fact that $(\rho,\tau)$ solves the system \eqref{sing_sys}, we obtain  
\[
R(z)-\tau=\Ord{\sqrt{1-z/\rho}},
\]
and using this information in the Taylor
expansion, the asymptotic relation
\begin{equation} \label{normal_R_asym}
R(z) \sim \tau -\sqrt\frac{2\rho F_z(\rho,\tau)}{F_{RR}(\rho,\tau)}\sqrt{1-\frac z\rho},
\text{ as } z\to \rho
\end{equation}
is easily deduced, provided that $F_{RR}(\rho,\tau)\neq 0$. Expanding the Taylor series further, $R(z)$ could be asymptotically expanded to
arbitrary order. These ideas are already found in \cite{Ev}. From \eqref{sing_sys} even exact,
however very involved, expressions for $\rho$ and $\tau$ can be found. We confine ourselves with
the numerical approximations
$$
\rho\approx 0.2382575749\dots \quad\text{ and }\quad \tau \approx 0.35829618282\dots.
$$
Thus,
$$
F_z(z,R)=1 \quad\text{ and }\quad F_{RR}(z,R)=1+\frac{3R}{(1-R)^4},
$$
and so $F_{RR}(\rho,\tau) \approx 7.3390612092\dots.$ This implies
$$
\sqrt\frac{2\rho F_z(\rho,\tau)}{F_{RR}(\rho,\tau)}\approx 0.2548109584\dots,
$$
from which we infer
\begin{equation} \label{r_n_normal}
\frac{r_n}{n!}=[z^n]R(z)\sim \frac{C_R}{\sqrt{\pi}n^{3/2}}\rho^{-n}, \text{ with }
C_R=\sqrt\frac{\rho F_z(\rho,\tau)}{2F_{RR}(\rho,\tau)}\approx 0.12740547924\dots
\end{equation}
after all. Moreover, from \eqref{normal_R_asym} and Remark~\ref{diff-sing-exp}, we obtain
\begin{equation} \label{R_p_normal}
R'(z)\sim \sqrt\frac{F_z(\rho,\tau)}{2\rho F_{RR}(\rho,\tau)}
\rcp{\sqrt{1-\frac z\rho}}, \text{ as } z\to \rho,
\text{ where }
\sqrt\frac{F_z(\rho,\tau)}{2\rho F_{RR}(\rho,\tau)}\approx 0.5347384202\dots.
\end{equation}

For the bivariate function $R(z,x)$ of $S_n^{(\ell,\max)}$, we get, arguing as in the previous sections, the functional
equation
\begin{align}
R(z,x)&= z+\rcp2 R(zx,x)^2+\sum_{\ell\ge 1}\sum_{k>\ell}\prod_{i=1}^{k+1} R(zx^{i+1},x) \prod_{j=1}^\ell R(zx^{j+1},x)
\nonumber \\
&\qquad+\rcp 2\sum_{k\ge 1} R(zx^{k+2},x)\prod_{i=1}^k R(zx^{i+1},x)^2.
\label{normal_biv}
\end{align}
As before, set $S(z):=R_x(z,1)$ and differentiate the functional equation with respect to $x$ and
let $x=1$ to get a linear equation for $S(z)$ having the solution
$$
S(z)=\frac{zR'(z)f(R(z))}{1-g(R(z))}
$$
with
$$
%f(t)=\frac{t(2t^7-4t^6+9t^4-17t^3-3t^2+27t+2)}{2(1+t)^3(1-t)^4} \quad\text{ and }\quad
f(t)=\frac{t(2t^5-4t^4-2t^3+5t^2+t+2))}{2(1+t)^3(1-t)^4} \quad\text{ and }\quad
g(t)=-\frac{t(2t^3-5t^2+3t-2)}{2(1-t)^3}.
$$
By means of \eqref{normal_R_asym} and \eqref{R_p_normal}, the singularity expansion of $S(z)$ is given by
\[
S(z)\sim \frac{\rho R'(\rho)s(\tau)}{1-\frac z\rho}, \text{ as } z\to\rho
\]
with
$$
%s(t)=\frac{t\(2t^7-4t^6+9t^4-17t^3-3t^2+27t+2\)}
s(t)=\frac{t\(2t^5-4t^4-2t^3+5t^2+t+2\)}
{2(1+t)^3 (1-t)^4
{\displaystyle\(\frac{6C_R}{1-t}-\frac{t (6C_R t^2 -10 C_R t +3 C_R)}{(1-t)3}+\frac{2C_R}{t}\)}}.
$$
Numerically, we get
%$C_S:=\rho R'(\rho)s(\tau) \approx 0.305844102\dots$
$C_S:=\rho R'(\rho)s(\tau) \approx 0.0819756013\dots$,
and so we get finally $[z^n] S(z) \sim C_S \rho^{-n}$. This in conjunction with \eqref{r_n_normal}
eventually gives
$$
\E{S_n^{(\ell,\max)}} \sim C^{+} \sqrt{\pi} n^{3/2} \quad \text{ with }\quad
%\frac{C_S}{C_R}\approx 2.4005569024\dots.
C^{+}=\frac{C_S}{C_R}\approx 0.6434228878\dots.
$$

Turning to $S_n^{(\ell,\min)}$, in the functional equation \eqref{normal_biv}, we have to change the
range of the two products in the double sum: Instead of $\prod_{i=1}^{k+1} \dots
\prod_{j=1}^\ell\dots$, we must have $\prod_{i=1}^{k} \dots \prod_{j=1}^{\ell+1}\dots$, which is
the only change. Similar computations as before yield then
$$
\E{S_n^{(\ell,\min)}} \sim C^- \sqrt{\pi} n^{3/2} \quad\text{ with }\quad C^-\approx
%2.2092954704\dots.
0.6062795760\dots.
$$

\section{Unlabeled Galled Trees}\label{unlabeled}

For the unlabelled counterparts of the considered leaf-labeled classes of galled trees, we again only consider the derivation of the
asymptotics of the mean (Theorem~\ref{main-thm-1}), in particular, the computation of the multplicative constants;
the limit law (Theorem~\ref{main-thm-2}) again follows with similar tools as in Section~\ref{ll} (only minor modifications are needed).

We use the same specifications \eqref{R_seteq}, \eqref{simplex_spec}, and \eqref{normal_spec}, respectively.
The only difference lies in the dictionary that is used to translate this into functional
equations for the generating functions, and within there only the treatment of the symmetric
combinatorial product differs from the labeled cases. In the following subsections, we will briefly
sketch how to deal with unlabeled classes.

\subsection{General case}

In the general case, we deal with unlabeled galled trees without any further constraint. They are
described in Figure~\ref{R_symb} and specified by \eqref{R_seteq}. Using the translation scheme
for unlabeled structures, this leads to the functional equation
\begin{equation}\label{R_unlab_funeq}
R(z)=z+\frac{R(z)^2+R(z^2)}2+\frac{R(z)^2}{1-R(z)}
+\frac{R(z)}2\(\(\frac{R(z)}{1-R(z)}\)^2+\frac{R(z^2)}{1-R(z^2)}\).
\end{equation}
Due to the terms involving $z^2$, it is impossible to derive an explicit formula for $R(z)$. We
will, however, again appeal to the implicit function theorem to get the dominant singularity $\rho$
and $\tau=R(\rho)$. Setting $\sigma=R(\rho^2)$, then we have (taking \eqref{R_unlab_funeq} at $z=\rho$ and its
derivative with respect to $\tau$)
\begin{align*}
\tau&=\rho+\frac{\tau^2+\sigma}2+\frac{\tau^2}{1-\tau}+
\frac{\tau}2\(\(\frac{\tau}{1-\tau}\)^2+\frac{\sigma}{1-\sigma}\), \\[5pt]
1&= \tau+\frac{2\tau-\tau^2}{(1-\tau)^2}+\frac{3\tau^2-\tau^3}{(1-\tau)^3}
+\rcp2\frac{\sigma}{1-\sigma}.
\end{align*}
Note further that, if we write \eqref{R_unlab_funeq} as $R(z)=\Psi(R(z))$, then $\Psi$ is an
operator on the set of formal power series that is a contraction with respect to the formal
topology (\cf \cite[Appendix A]{FlSe}). If we start from the null series and iterate $\Psi$, then
the iterate $\Psi^m(0)$ is a power series that coincides in its first $m$ terms with $R(z)$. As
$\rho<1$, for $0<z<\rho$, we have $z^2< z$. This implies that $R(z^2)$ converges very fast,
which enables us to get accurate approximations of $\sigma=R(\rho^2)$. Using this value, the system above
can be solved numerically for $\rho$ and $\tau$. We get
$$
\rho\approx 0.1164742652\dots \quad\text{ and }\quad \tau\approx 0.2182060577\dots,
$$
and also $\sigma\approx 0.0139559001\dots$.
\brem
Caveat: Approximating $\tau$ by solving for $\rho$ and inserting into $R(z)$ would be very
inefficient, as $R(z)$ converges very slowly at $z=\rho$. Indeed, the convergence rate is only
$\Theta(1/\sqrt n)$ if $n$ terms of the power series are used.
\erem
As in the labeled cases, expanding the functional equation $F(z,R(z))=0$ at the singularity gives the singular expansion \eqref{normal_R_asym}, where $F$ is given by
\[
F(z,R)=-R+z+\frac{R^2+\sigma}2+\frac{R^2}{1-R}
+\frac{R}2\(\(\frac{R}{1-R}\)^2+\frac{\sigma}{1-\sigma}\).
\]
Taking partial derivatives gives
\begin{align*}
F_{RR}(z,R)&=\frac{R^4-4R^3+6R^2-3R+3}{(1-R)^4}, \\
F_z(z,R)&= 1+zR'(z^2)+\frac{zR'(z^2)R}{1-R(z^2)},
\end{align*}
and so we obtain
$$
R(z)\sim \tau-C\sqrt{1-\frac z\rho}\quad\text{ where }\; C\approx  0.196590883\dots.
$$
This implies
$$
\frac{r_n}{n!}=[z^n]R(z)\sim\frac{C_R}{\sqrt{\pi n^3}}\rho^{-n}\quad\text{with}\quad C_R=C/2.
$$

Now, let us turn to the Sackin index, where we again start with $S^{+}$. Like in the labeled cases, we obtain from
\eqref{R_unlab_funeq} the functional equation for $S(z)=R_x(z,x)$, with $x$ keeping track of
the value of the Sackin index. By the additivity of the Sackin index, the treatment of the symmetric
combinatorial product for the multivariate function $R(z,x)$ is like for $R(z)$. For $R(z,x)$, we
get
\begin{align*}
R(z,x)=z&+\frac{R(zx,x)^2+R(z^2x^2,x^2)}2
+\sum_{\ell\ge 0}\sum_{k>\ell} \prod_{i=1}^{k+1}R(zx^{i+1},x) \prod_{j=1}^\ell R(zx^{j+1},x)\\
&+\rcp2\sum_{k\ge1} R(zx^{k+2},x) \left(\prod_{i=1}^k R(zx^{i+1},x)^2+\prod_{i=1}^k R(z^2x^{2i+2},x^2)\right),
\end{align*}
and for $S(z)$ this implies
$$
S(z)=\frac{zR'(z)f(z,R(z))+h(z,R(z))}{1-g(z,R(z))}
$$
with
\begin{align*}
f(z,t)&=
%\frac{t(2t^7-4t^6+9t^4-17t^3-3t^2+27t+2}{2(1+t)^3(1-t)^4}
%\frac{(2t^8-4t^7-4t^6+13t^5-3t^4-15t^3+7t^2+10t+10}{2(1+t)^3(1-t)^4}
\frac{2t^5-8t^4+10t^3-t^2-11t+12}{2(1+t)^3(1-t)^4}
+\frac{R(z^2)(3-2R(z^2))}{2(1-R(z^2))^2},\\
%g(t)&=\frac{t(-2t^3+5t^2-3t+2)}{2(1-t)^3}+\frac{R(z^2)}{2(1-R(z^2))} \\
g(z,t)&=\frac{t(-2t^3+7t^2-9t+6)}{2(1-t)^3}+\frac{R(z^2)}{2(1-R(z^2))}, \\
h(z,t)&=z^2 R'(z^2)\(1+\frac{t(2-R(z^2))}{(1-R(z^2))^3}\) + S(z^2)\(1 +\frac{t}{(1-R(z^2))^2}\).
\end{align*}
The next step is to determine the singular expansion of $S(z)$, namely,
$$
S(z)\sim \frac{C_S}{1-\frac z\rho}, \quad\text{ as } z\to\rho,
$$
Luckily enough,\footnote{The reason for this is that the singularity stems from the denominator $1-g(z,R(z))$ and from $R'(z)$. The latter is a factor of $f(z,R(z)$ but not of $h(z,R(z)$ and hence makes the singularity of the first term more dominant. Therefore, the  nasty terms contribute only to lower order terms of the asymptotic expansion.} the nasty terms $S(z^2)$ and $R'(z^2)$, which would require a cumbersome numerical
treatment do not show up in the main coefficient $C_S$ which happens to depend only on $\rho$,
$\tau$, $\sigma$, and $C$. So, we easily compute that $C_S\approx
0.168075882\dots$ and
obtain finally
\[
\E{S_n^{(u,\max)}} \sim C^{+} \sqrt{\pi} n^{3/2} \quad \text{ with }\quad
C^{+}=\frac{C_S}{C_R}\approx
1.7099051570\dots,
\]
and in a similar way for the other variant of the Sackin index:
\[
\E{S_n^{(u,\min)}} \sim C^- \sqrt{\pi} n^{3/2} \quad \text{ with }\quad
C^-\approx
1.4350664453\dots.
\]

\subsection{Simplex galled trees}

As before, we start from the generic specification \eqref{simplex_spec} but take the unlabeled
translation scheme instead. We obtain
\[
R(z)=z+\frac{R(z)^2+R(z^2)}2+\frac{zR(z)}{1-R(z)}
+\frac{z}2\(\(\frac{R(z)}{1-R(z)}\)^2+\frac{R(z^2)}{1-R(z^2)}\).
\]
Like in the general case, we are faced with a functional equation of the form $R(z)=\Psi(R(z))$
with $\Psi$ being a contraction on the set of formal power series. Again, $R(\rho^2)$ converges
rapidly and so the corresponding system for $\rho$ and $\tau=R(\rho)$, namely 
\begin{align*}
R(z)=\Psi(R(z)), \qquad 1=\Psi'(R(z)),     
\end{align*}
can be easily solved numerically. This gives
$$
\rho=0.162165279\dots \quad\text{ and }\quad \tau=0.365415487\dots.
$$
This permits the numerical computation of the singular expansion \eqref{normal_R_asym} of $R(z)$,
thus giving us hands on the asymptotics of its coefficients. Here, the multiplicative scaling
factor of the singular term in $R(z)\sim \tau -C\sqrt{1-\frac z\rho}$ is
$C\approx 0.3995272519\dots$, which gives the coefficient asymptotics $[z^n]R(z)\sim
\(C_R/\sqrt{\pi n^3}\)\rho^{-n}$ with $C_R=C/2$.

Likewise, the bivariate problem is treated as in the previous section. The functional equation for
$R(z,x)$ of $S^{+}$ reads as
\begin{align}
R(z,x)=z&+\frac{R(zx,x)^2+R(z^2x^2,x^2)}2+ \sum_{\ell\ge 0}\sum_{k>\ell} zx^{k+2}
\prod_{i=1}^{k}  R(zx^{i+1},x) \prod_{j=1}^\ell R(zx^{j+1},x)\nonumber\\
&+\rcp2\sum_{k\ge1} zx^{k+2} \left(\prod_{i=1}^k R(zx^{i+1},x)^2+\prod_{i=1}^k R(z^2x^{2i+2},x^2)\right), \label{multivar_funeq}
\end{align}
implying
\begin{equation} \label{S_simplex_unlab}
S(z)=\frac{zR'(z)f(z,R(z))+h(z,R(z))}{1-g(z,R(z))}
\end{equation}
with
$$
f(z,t)=t+z\frac{2-t}{(1-t)^4},\qquad g(z,t)=t+\frac{z}{(1-t)^3},
$$
and
\begin{align}
h(z,t)&=z^2 R'(z^2)\(1+\frac{z(2-R(z^2))}{(1-R(z^2))^3}\)
+S(z^2)\(1 +\frac{z}{(1-R(z^2))^2}\) \nonumber \\[2mm]
& \quad
+z\frac{3R(z^2)-2R(z^2)^2}{(1-R(z^2))^2}
+z\frac{t(t^3-4t^2-t+6)}{2(1-t)^3 (1+t)}. \label{h_simplex_unlab}
\end{align}

Now use the singular expansion of $R(z)$ at $\rho$ to obtain that $S(z)\sim C_S\(1-\frac
z\rho\)^{-1}$, as $z\to\rho$, where $C_S=1/4$. Eventually, this gives
\[
\E{S_n^{(u,\max)}} \sim C^{+}\sqrt{\pi} n^{3/2} \quad \text{ with }\quad
C^{+}=\frac{C_S}{C_R}\approx
1.2514790858\dots,
\]
and similarly
\[
\E{S_n^{(u,\min)}} \sim C^- \sqrt{\pi} n^{3/2} \quad \text{ with }\quad
C^-=C^+.
\]

As in the labeled case for simplex trees, we encounter here the phenomenon that the two variants
of the Sackin index are asympotically the same. Their difference shows up only in the
second-order term of the asymptotics.

This time, we do not have an explicit expression for $R(z)$ that can be expanded, as in the
labeled case. So, we need to go back to the functional equation and expand $F(z,R)$ further into a
Taylor series at $(\rho,\tau)$, noting that $F(z,R(z))=0$, $F_R(\rho,\tau)=0$ and that
$R(z)-\tau\sim -C\sqrt{z-\rho}$, as $z$ approaches $\rho$, where $C=\sqrt{\rho
F_z(\rho,\tau)/F_{RR}(\rho,\tau)}$, \emph{cf.} \eqref{normal_R_asym}. This yields, as $z\to\rho$,
\[
0=F_z(z-\rho)+\frac{F_{RR}}2 (R-\tau)^2+ F_{zR}
(z-\rho)(R-\tau)+\frac{F_{RRR}}6(R-\tau)^3+\Ord{|z-\rho|^2},
\]
where the partial derivatives of $F$ are evaluated at $(\rho,\tau)$. Consequently,
\begin{align*}
R(z)&\sim \tau -C \sqrt{1-\frac z\rho} \sqrt{1+C\cdot\(\frac{F_{RRR}}{3F_{RR}}
-\frac{F_{zR}}{F_z}\)} \\
&\sim \tau -C \sqrt{1-\frac z\rho} -\frac{C^2}2\(\frac{F_{RRR}}{3F_{RR}}
-\frac{F_{zR}}{F_z}\) \(1-\frac z\rho\) \\
&=\tau -C \sqrt{1-\frac z\rho}+D\(1-\frac z\rho\)
\end{align*}
with $D\approx 0.0520951948$, which implies, by Remark~\ref{diff-sing-exp},
\[
R'(z)\sim \frac{C}{2\rho}\(1-\frac z\rho\)^{-1/2}-\frac D\rho.
\]
This enables us to obtain the second-order term in the singular expansions of $f(z,R(z))$ and
$g(z,R(z))$:
\[
f(z,R(z))\sim 2+D_f\sqrt{1-\frac z\rho}\ \ \text{and}\ \ 1-g(z,R(z))\sim C_g\sqrt{1-\frac z\rho}+
D_g\(1-\frac z\rho\)
\]
with $D_f\approx -4.1164638932\dots$, $C_g\approx 1.5981090076\dots$ and $D_g\approx
-1.5092271101\dots$. From \eqref{S_simplex_unlab}, we get then
\[
S(z)\sim \frac{C_S}{1-\frac z\rho} + \frac{C D_f C_g - 2D C_f C_g - C C_f D_g+2C_gH}{2C_g^2}
\rcp{\sqrt{1-\frac z\rho}},
\]
where $H=\lim_{z\to\rho} h(z,R(z))$. The computation of $H$ requires the value of $S(\rho^2)$.

One way to do this is appealing to the functional equation~\eqref{multivar_funeq} and
using the fact that the right-hand side can be seen as the application of an operator to $R(z,x)$
that is a contraction in the formal metric. The convergence is exponential, so not many
coefficients are needed, but as the coefficient $z^n$ is a polynomial in $x$ and we get only one
additional correct coefficient per iteration, this requires heavy computations.

Alternatively, we may use \eqref{S_simplex_unlab}. The function $S(z)$ depends on $h(z,R(z))$ which in turn depends on $S(z^2)$. Thus, applying \eqref{S_simplex_unlab} iteratively this means that we actually insert
values of the form $z^{2^i}$ into all the involved functions, which proves very efficient. We obtain $S(\rho^2)\approx 0.0059339813\dots$, and can therefore compute $H$.
Here also lies the only difference between the two Sackin indices. Computing $H$ for both
cases gives the desired result after all:
\[
S^{\pm}(z)\sim \frac{C_S}{1-\frac z\rho}+\frac{D_S^{\pm}}{\sqrt{1-\frac z\rho}}
\]
with $D_S^+\approx -0.033882959\dots$ and $D_S^-\approx -0.140151617\dots$, which leads to the
final result
\begin{align*}
\E{S_n^{(u,\max)}} &\sim K\sqrt{\pi} n^{3/2} + \frac{C_R}{D_S^+} n,
\\
\E{S_n^{(u,\min)}} &\sim K\sqrt{\pi} n^{3/2} + \frac{C_R}{D_S^-} n,
\end{align*}
where $K=C^+=C^-$ and $C_R/D_S^+\approx -0.1696152609\dots$ and
$C_R/D_S^-\approx-0.701587271\dots$.

\subsection{Normal galled trees}

In the case of normal galled trees, we play the same game again. In the specification
\eqref{normal_spec} as before, we get the functional equation
\[
R(z)=z+\frac{R(z)^2+R(z^2)}2+\frac{R(z)}2\(\(\frac{R(z)}{1-R(z)}\)^2+\frac{R(z^2)}{1-R(z^2)}\).
\]
From this, we obtain in the way as in the other cases the numerical values of the crucial
constants:
\[
\rho=0.207339752\dots \quad\text{ and }\quad \tau=0.35504356\dots.
\]

Extending to the bivariate function of $S^{+}$ leads to
\begin{align*}
R(z,x)=z&+\frac{R(zx,x)^2+R(z^2x^2,x^2)}2+ \sum_{\ell\ge 1}\sum_{k>\ell} \prod_{i=1}^{k+1}  R(zx^{i+1},x) \prod_{j=1}^\ell R(zx^{j+1},x)\\
&+\rcp2\sum_{k\ge1} R(zx^{k+2},x)\left(\prod_{i=1}^k R(zx^{i+1},x)^2+\prod_{i=1}^k R(z^2x^{2i+2},x^2)\right),
\end{align*}
and after the differentiation process this gives
$$
S(z)=\frac{zR'(z)f(z,R(z))+h(z,R(z))}{1-g(z,R(z))}
$$
with
\begin{align*}
f(z,t)&=
\frac{2t^5-4t^4-2t^3+5t^2+t+2}{2(1+t)(1-t)^4}
+\frac{R(z^2)(3-2R(z^2))}{2(1-R(z^2))^2}, \\
g(z,t)&=\frac{t(-2t^3+5t^2-3t+2)}{2(1-t)^3}+\frac{R(z^2)}{2(1-R(z^2))}, \\
h(z,t)&=z^2 R'(z^2)\(1+\frac{t(2-R(z^2))}{(1-R(z^2))^3}\) + S(z^2)\(1 +\frac{t}{(1-R(z^2))^2}\).
\end{align*}

Again, the singular expansion \eqref{normal_R_asym} of $R(z)$ is the clue to get the coefficient
asymptotics for $R(z)$ and $S(z)$, which gives the asymptotic expression for the Sackin index
after all:
\[
\E{S_n^{(u,\max)}} \sim C^+ \sqrt{\pi} n^{3/2} \quad \text{ with }\quad
C^+\approx
1.125542584\dots.
\]
Similarly, we get
\[
\E{S_n^{(u,\min)}} \sim C^- \sqrt{\pi} n^{3/2} \quad \text{ with }\quad
C^-\approx
1.0632588514\dots.
\]

\section{Conclusion}\label{con}

We first summarize the contributions of this paper. We proposed two extensions of the Sackin index to phylogenetic networks, namely, the Sackin index where the longest (resp. the shortest) path to a leaf is always chosen. The former was already considered in \cite{Zh} where the order of the mean was derived for random simplex labeled tree-child networks which are sampled uniformly at random from the set of all simplex tree-child networks with $n$ leaves. In this work, we derived the first-order asymptotics of the mean for both indices for random galled trees and variants again under the uniform random model; in addition, we considered both the labeled and unlabeled case. Moreover, we extended our results to all higher moments and also showed that the (proper normalized) Sackin indices converge (weakly and with all their moments) to the Airy distribution. Thus, the Sackin index of a random galled tree behaves similar as the Sackin index of phylogenetic trees under the PDA model; see \cite{BlFrJa}.

One question, which was left open by our study, is which Sackin index is better suited to measure the ``balance" of a phylogenetic network? Or should a totally different extension of the Sackin index be used? In order to answer these questions, a combinatorial study similar to \cite{KnFiHeWi} has to be performed. The latter paper answers these questions for the extension of another popular balance index for phylogenetic trees, namely, the total cophenetic index. Note, however, that \cite{KnFiHeWi} does not study stochastic properties of the proposed extension.

Yet another extension, which by its very definition does measure the balance of a phylogenetic network, is the $B_2$ index for which a similar study as in the current paper will be performed in the companion paper \cite{BiDuFuYu}. Again, asymptotics for all moments and a limit distribution result for galled trees will be derived (with a combinatorial approach similar to the one used in the current paper and a probabilistic approach based on local limits).

We conclude by pointing out that the current paper also solves a couple of (so far unsolved) asymptotic counting problems for classes of galled trees. More precisely, for the general labeled class of galled trees, such a result was already presented in \cite{BoGaMa}. However, simplex and normal labeled galled trees have only been counted exactly in \cite{CaZh}. The expressions from this paper give little insight into the asymptotics of these numbers which were derived in the current paper. As for unlabeled classes, only unlabeled normal galled trees have been considered so far. In \cite{AgMaRo}, the authors derived an asymptotic counting result but with a different approach which in particular did not use symbolic combinatorics which is the method used here. This approach allowed a more compact derivation of the first-order asymptotics as well as the straightforward extensions to variants.

\section*{Data Availability Statement} 

No datasets were generated or analyzed during the current study.

\section*{Acknowledgment} 

The authors express their gratitude to two anonymous referees who carefully read the  manuscript and pointed out several errors and provided numerous suggestions improving the presentation. In particular, we thank the referee who checked the numerical computations and reported an error in the second-order asymptotics of simplex labeled galled trees.

\end{document}